\documentstyle[12pt,rotate]{article}

\input epsf.sty

\newcommand{\insertfig}[2]{\mbox{\epsfxsize=#1cm \epsfbox{#2.eps}}}

\newcommand{\Bx}{x_{\rm B}}
\newcommand{\ft}[2]{{\textstyle\frac{#1}{#2}}}
\newcommand{\MS}{{$\overline{\mbox{MS}}$}}

\begin{document}

\begin{titlepage}

\centerline{\large\bf Hard exclusive meson production at next-to-leading
                      order.}

\vspace{15mm}

\centerline{\bf A.V. Belitsky$^{a}$, D. M\"uller$^{b,c,a}$}

\vspace{10mm}

\centerline{\it $^a$C.N.\ Yang Institute for Theoretical Physics}
\centerline{\it State University of New York at Stony Brook}
\centerline{\it NY 11794-3840, Stony Brook, USA}

\vspace{5mm}

\centerline{\it $^b$Fachbereich Physik, Universit\"at Wuppertal}
\centerline{\it D-42097 Wuppertal, Germany}

\vspace{5mm}

\centerline{\it $^c$Institut f\"ur Theoretische Physik,
                Universit\"at Regensburg}
\centerline{\it D-93040 Regensburg, Germany}

\vspace{2cm}

\centerline{\bf Abstract}

\vspace{0.5cm}

\noindent
We evaluate perturbative next-to-leading order corrections to the hard
exclusive leptoproduction of $\pi^+$ mesons on a transversely polarized
proton target. A model dependent study shows that these corrections can 
be large. We analyze the scale dependence and explore the
Brodsky-Lepage-Mackenzie scale setting procedure. Although the predictions 
for the cross section suffer from theoretical uncertainties, the 
transverse nucleon single spin asymmetry turns out to be a rather stable 
observable since higher order effects approximately cancel there.

\vspace{4.5cm}


\vspace{0.5cm}


\end{titlepage}

%%%%%%%%%%%%%%%%%%%%%%%%%%%%%%%%%%%%%%%%%%%%%%%%%%%%%%%%%%%%%%%%%%%%%
\section{Exclusive meson production and QCD factorization.}
\label{Sec-Int}
%%%%%%%%%%%%%%%%%%%%%%%%%%%%%%%%%%%%%%%%%%%%%%%%%%%%%%%%%%%%%%%%%%%%%

\begin{figure}[t]
\begin{center}
\mbox{
\begin{picture}(0,85)(100,0)
\put(-65,-14){\insertfig{11}{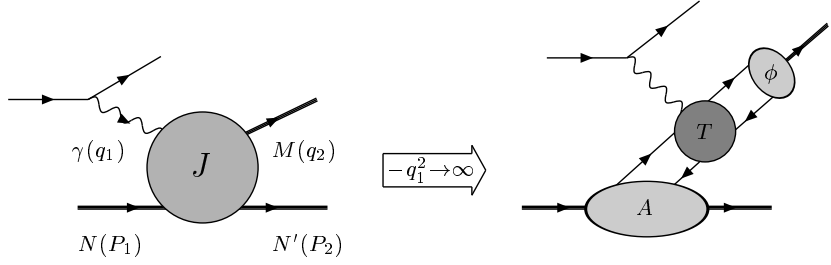}}
\end{picture}
}
\end{center}
\caption{\label{Factorization} Factorization of the meson leptoproduction at
$- q_1^2 \to \infty$ into hard-scattering amplitude $T$ and non-perturbative
functions $A$, the generalized flavour changing parton distribution, and $\phi$,
the distribution amplitude of the outgoing meson.}
\end{figure}

The hard exclusive leptoproduction of a meson $M$ from a nucleon target
$N$,
\begin{equation}
\ell (k) N (P_1) \to \ell' (k') N' (P_2) M (q_2)
\end{equation}
is a promising process to test our understanding of QCD in exclusive
reactions, as well as a means for studies of the properties of nucleon
to hadron transitions, $N \to N'$, with $N'$ being a baryon from an
$SU_f (3)$ multiplet. If the intermediate photon is longitudinally
polarized and has a large virtuality ${\cal Q}^2 = - q_1^2$, the
photoproduction amplitude $\gamma_{\scriptscriptstyle L} N \to N' M$ factorizes
into a convolution of three parts \cite{ColFraStr96}, see
Fig.\ \ref{Factorization},
\begin{eqnarray}
\label{LongAmpl}
{\cal A}
= \sum_{f,f^\prime = u, d, s}
\int_0^{1} d u \int_{-1}^{1} d x \,
\phi_{f f^\prime} (u | \mu)
T_{f f^\prime} (u, x, \xi | {\cal Q}, \mu)
A_{f f^\prime} (x, \xi , \Delta^2 | \mu) + \dots,
\end{eqnarray}
where the ellipsis stand for power suppressed contributions in $1/{\cal Q}$.
The hard subprocess $T_{f f^\prime}$ encodes the short distance dynamics
of the parton scattering and can be consistently calculated in QCD 
perturbation theory as a series in the strong coupling $\alpha_s$. The other
two blocks, $\phi$ and $A$, are universal, i.e.\ process independent, and
embody the long-distance physics. $\phi$ is a conventional leading twist
meson distribution amplitude. It describes the minimal Fock component of the 
meson wave function with two quarks having momentum fractions $u$ and $1 - u$
w.r.t.\ the meson momentum, respectively. $A$ is a flavour nondiagonal
generalized parton distribution (GPD). Its definition, taking apart the
flavour transition, differs from the conventional Feynman parton densities
by the presence of a non-zero momentum flow, $\Delta = P_2 - P_1 = q_1 - q_2$,
in the $t$-channel. As a result, the GPD is a complicated function of
three variables, the $s$-channel momentum fraction $x$, its $t$-channel
counterpart\footnote{Note that in the kinematics, we are considering, the
skewedness parameter $\eta$ is approximately equal to a (negative) generalized
Bjorken variable $\xi$, to be specified below. } $\eta \sim \Delta_+$, and
$\Delta^2$. The appearing phase space picture is rich and results into a
trinity interpretation of GPDs: they share common properties with forward
parton densities, distribution amplitudes, and form factors.

%%%%%%%%%%%%%%%%%%%%%%%%%%%%%%%%%%%%%%%%%%%%%%%%%%%%%%%%%%%%%%%%%%%%%
%            Figure 1
%%%%%%%%%%%%%%%%%%%%%%%%%%%%%%%%%%%%%%%%%%%%%%%%%%%%%%%%%%%%%%%%%%%%%
%%%%%%%%%%%%%%%%%%%%%%%%%%%%%%%%%%%%%%%%%%%%%%%%%%%%%%%%%%%%%%%%%%%%%

In the present study we will concentrate on the next-to-leading order (NLO)
analysis of the $\pi^+$-production from the proton. Thus we set $N = p$,
$N' = n$ and $M = \pi^+$ in Fig.\ \ref{Factorization}. This reaction is an
issue of intensive experimental studies by HERMES \cite{Ste01} and Jefferson
Lab \cite{Sab01} in relation to the measurement of the helicity-flip GPDs
accessible in this process. Our consequent presentation is organized as
follows. In the next section we introduce the definitions of all basic
ingredients entering the amplitude (\ref{LongAmpl}) and calculate the
electroproduction cross section for the transversely polarized proton. In
section \ref{NLOanalytical} we present one-loop corrections for the
hard-scattering amplitude. In section \ref{ScaleSet} we introduce simple models
for the GPDs and then we discuss the issue of scale setting. After this
we give numerical estimates of the differential cross section and transverse
single spin asymmetry. Finally, we summarize.

%%%%%%%%%%%%%%%%%%%%%%%%%%%%%%%%%%%%%%%%%%%%%%%%%%%%%%%%%%%%%%%%%%%%%
\section{Amplitude and cross section for
$\gamma_{\scriptscriptstyle L} p \to \pi^+ n$.}
\label{HadronicAmplitude}
%%%%%%%%%%%%%%%%%%%%%%%%%%%%%%%%%%%%%%%%%%%%%%%%%%%%%%%%%%%%%%%%%%%%%

The hadronic part of the $\gamma p \to \pi^+ n$
process is given by the Fourier transform of the matrix element of the
electromagnetic current $J_\mu = \sqrt{4 \pi \alpha_{\rm em}} \, \sum_i
Q_i \bar \psi_i \gamma_\mu \psi_i$
\begin{equation}
\label{CurrentToAmplitude}
\int d^4 x {\rm e}^{- i q_1 \cdot x}
\langle \pi^+ (q_2) n (P_2) |
J_\mu (x)
| p (P_1) \rangle
= i (2 \pi)^4 \delta^{(4)} \left( q_1 + P_1 - q_2 - P_2 \right)
{\cal A}_\mu^{\pi^+} ,
\end{equation}
with $q_1 = k - k'$ being the difference of incoming and outgoing 
lepton momenta. Its leading term in $1/{\cal Q}^2$ is picked up by 
the contraction  with the longitudinal polarization vector 
$\varepsilon_{\scriptscriptstyle \! L}$. A straightforward leading 
twist calculation gives for the amplitude
\begin{equation}
\label{Amplitude}
{\cal A}_\mu^{\pi^+} = \sqrt{4 \pi \alpha_{\rm em}} \,
\frac{\pi}{N_c} \frac{f_\pi}{q^2} j_\mu
\int_{0}^{1} d u \int_{-1}^{1} d x \, \phi_\pi (u)
T_{ud} \left( u, x, \xi \right)
\frac{q \cdot A^{ud}}{q \cdot P} (x, \xi, \Delta^2) +
{\cal O}\left(1/q^3\right) ,
\end{equation}
where $q \equiv \ft12 \left( q_1 + q_2 \right)$, $P = P_1 + P_2$ and $\Delta
= q_1 - q_2 = P_2 - P_1$. The scaling variable is a generalized Bjorken
variable $\xi = - \frac{q^2}{q \cdot P} \approx - \frac{q \cdot \Delta}{q
\cdot P}$. Here in the last equality we neglected the pion mass, $m_\pi = 0$,
and power suppressed corrections in the nucleon mass $M^2/(- q^2)$. The
Lorentz structure $j_\mu = 2 q_\mu + 3 \xi P_\mu$, appearing in the
twist-two part of ${\cal A}_\mu^{\pi^+}$, is gauge invariant to the
twist-four accuracy, i.e.\ $q_{1 \mu} {\cal A}_\mu^{\pi^+} = \left( q +
\frac{\Delta}{2} \right)_\mu {\cal A}_\mu^{\pi^+} \approx 0$.

We introduced in Eq.\ (\ref{Amplitude}) two non-perturbative objects, the
pion distribution amplitude
\begin{equation}
\label{pionDA}
\langle \pi^+ (q_2) |
\bar u (y) \gamma_\rho \gamma_5 d (z)
| 0 \rangle
= - i q_{2 \rho} f_\pi \int_{0}^{1} d u \,
{\rm e}^{i u q_2 \cdot y + i (1 - u) q_2 \cdot z }
\phi_\pi (u) ,
\end{equation}
with the decay constant $f_\pi \approx 132 \ {\rm MeV}$, as well as
$A^{ud}_\rho$ expressed in terms of the off-forward
matrix element of the non-local flavour-changing operator
\begin{equation}
\label{GPD}
\langle n (P_2) |
\bar d (y) \gamma_\rho \gamma_5 u (z)
| p (P_1) \rangle
=
\int_{-1}^{1} d x \,
{\rm e}^{\frac{i}{2} \left( \Delta + x P \right) \cdot y
+ \frac{i}{2} \left( \Delta - x P \right) \cdot z }
A^{ud}_\rho (x, \xi, \Delta^2). 
\end{equation}
It gives rise to flavour non-diagonal GPDs, which enter as form factors
\begin{equation}
\label{FormFactorsGPD}
A^{ud}_\rho (x, \xi, \Delta^2)
= \widetilde h_\rho \widetilde H^{ud} (x, \xi, \Delta^2)
+ \widetilde e_\rho \widetilde E^{ud} (x, \xi, \Delta^2) ,
\end{equation}
in front of the Dirac structures
\begin{equation}
\label{DiracStructures}
\widetilde h_\rho
= \bar U_n (P_2) \gamma_\rho \gamma_5 U_p (P_1) ,
\qquad
\widetilde e_\rho
= \frac{\Delta_\rho}{2 M} \bar U_n (P_2) \gamma_5 U_p (P_1) .
\end{equation}
We can safely neglect the mass difference of proton and neutron and set in
the following $M \equiv M_p = M_n$. Using the isospin symmetry of strong
interactions, one can reduce the above GPDs to the conventional flavour
diagonal proton matrix elements \cite{ManPilWei97}
\begin{equation}
\label{SU2symm}
A^{ud}_\rho = A^{uu}_\rho - A^{dd}_\rho .
\end{equation}

%%%%%%%%%%%%%%%%%%%%%%%%%%%%%%%%%%%%%%%%%%%%%%%%%%%%%%%%%%%%%%%%%%%%%
%            Figure 2
%%%%%%%%%%%%%%%%%%%%%%%%%%%%%%%%%%%%%%%%%%%%%%%%%%%%%%%%%%%%%%%%%%%%%
\begin{figure}[t]
\begin{center}
\mbox{
\begin{picture}(0,38)(100,0)
\put(-120,-14){\insertfig{15}{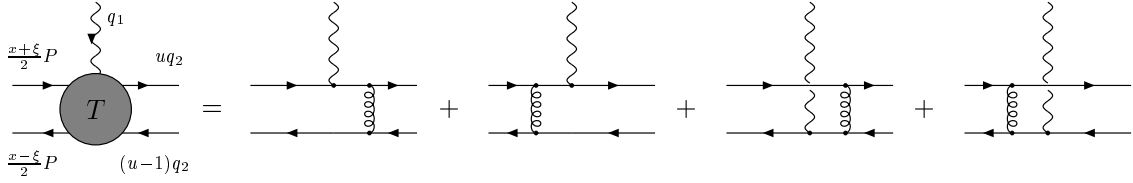}}
\end{picture}
}
\end{center}
\caption{\label{coefficient} Hard-scattering coefficient function at
leading order.}
\end{figure}
%%%%%%%%%%%%%%%%%%%%%%%%%%%%%%%%%%%%%%%%%%%%%%%%%%%%%%%%%%%%%%%%%%%%%

Evaluating the tree diagrams presented in Fig.\ \ref{coefficient}, we get
the known result for the hard amplitude $T_{ud}$ to leading order (LO)
accuracy in the QCD coupling \cite{ManPilRad99,FraPobPolStr99,VanGuiGui99},
\begin{equation}
\label{LOamplitude}
T_{ud} (u, x, \xi)
= C_F \alpha_s \frac{1}{\xi}
\left\{
\frac{Q_u}{(1 - u) \left( 1 - \ft{x}{\xi} - i 0 \right)}
-
\frac{Q_d}{u \left( 1 + \ft{x}{\xi} - i 0 \right)}
\right\}
+
{\cal O} (\alpha_s^2) .
\end{equation}
Here the quark charges are $Q_u = 2/3$ and $Q_d = -1/3$.

Let us remark on similarities of the present consideration to the one of
the pion form factor since we will use them below. Contributing diagrams can
be decomposed into two sets, where in the first (second) one the photon is
attached to the (anti-) quark line. The momentum of the initial and final
(anti-) quark is given in the collinear approximation by $p_1 = - \frac{x +
\xi}{2\xi} \Delta$ $\left( p'_1 = - \frac{x - \xi}{2\xi} \Delta \right)$ and
$p_2 = u q_2$ $\left( p'_2 = (1 - u) q_2 \right)$, respectively. In leading
twist approximation both of these sets separately respect current
conservation. Obviously, if we formally replace $\frac{x + \xi}{2\xi} \to v$
(then $\frac{x - \xi}{2\xi} \to 1 - v$) with $0 < v < 1$, we reduce the 
result (\ref{LOamplitude}) to the kinematics of the pion form factor.
Of course, we can also recover the amplitude (\ref{LOamplitude}) from the pion
form factor result by the substitution  $v\to \frac{x + \xi}{2\xi}$. In
addition we have to keep track of the emerging
imaginary part, which is absent in the pion form factor. It develops now in
the region $|x| \ge \xi$ and can  be easily restored. Using this analogy, the
NLO corrections of $T_{ud}$ will be evaluated in section \ref{NLOanalytical}.

Now we turn to the calculation of the cross section for electroproduction of
pions from the proton target
$\ell (k) p (P_1) \to \ell (k') n (P_2) \pi^+ (q_2)$. It is given by
\begin{equation}
\label{XsectionInv}
d \sigma^{\pi^+} = \frac{1}{4 k \cdot P_1}
\left|
L_\mu {\cal A}_\mu^{\pi^+}
\right|^2 d {\rm LIPS}_3 ,
\end{equation}
where the hadronic amplitude (\ref{Amplitude}) is contracted with the
leptonic current (which includes the photon propagator)
\begin{equation}
\label{LeptonCurrent}
L_\mu = \frac{i}{{\cal Q}^2}
\sqrt{4 \pi \alpha_{\rm em}} \, \bar u (k') \gamma_\mu u (k) ,
\end{equation}
and the three-particle phase space volume reads
\begin{equation}
\label{LIPS}
d {\rm LIPS}_3 = (2 \pi)^4 \delta^{(4)} (k + P_1 - k' - P_2 - q_2)
\frac{d^3 k'}{2 E' (2 \pi)^3}
\frac{d^3 P_2}{2 E_2 (2 \pi)^3}
\frac{d^3 q_2}{2 \varepsilon_2 (2 \pi)^3} .
\end{equation}
%%%%%%%%%%%%%%%%%%%%%%%%%%%%%%%%%%%%%%%%%%%%%%%%%%%%%%%%%%%%%%%%%%%%%
%            Figure 3
%%%%%%%%%%%%%%%%%%%%%%%%%%%%%%%%%%%%%%%%%%%%%%%%%%%%%%%%%%%%%%%%%%%%%
\begin{figure}[t]
\begin{center}
\mbox{
\begin{picture}(0,127)(100,0)
\put(-65,-14){\insertfig{11}{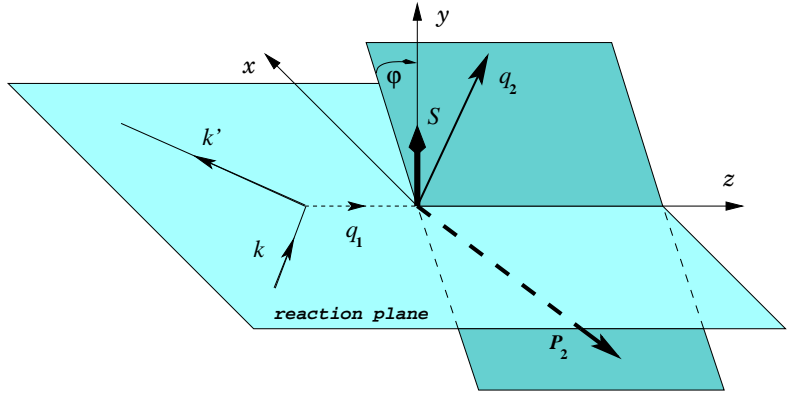}}
\end{picture}
}
\end{center}
\caption{\label{Kinemat} The reference frame for the cross section is
defined as a target rest-frame and the $z$-axis directed along
the three-vector of the virtual photon. The outgoing hadrons, pion and
neutron, are in the same plane (due to the momentum conservation), which
form an angle $\varphi$ with the transverse polarization vector $\vec S$
of the target.}
\end{figure}
%%%%%%%%%%%%%%%%%%%%%%%%%%%%%%%%%%%%%%%%%%%%%%%%%%%%%%%%%%%%%%%%%%%%%
In the rest-frame of the target with the $z$-axis chosen along the
momentum of virtual photon, as shown in Fig. \ref{Kinemat}, we get the
following four-fold cross section
\begin{equation}
\label{4Xsection}
\frac{d \sigma^{\pi^+}}{d {\cal Q}^2 d \Bx d |\Delta^2| d \varphi}
= \frac{1}{2 (4 \pi)^4} \frac{\Bx y^2}{{\cal Q}^4}
\left( 1 + 4 \frac{M^2 \Bx^2}{{\cal Q}^2} \right)^{-1/2}
\left|
L_\mu {\cal A}_\mu^{\pi^+}
\right|^2 ,
\end{equation}
where we use the variables ${\cal Q}^2 = - q_1^2$, $\Bx = - q_1^2/( 2
q_1 \cdot P_1)$, $y = q_1 \cdot P_1/ k \cdot P_1$, and $\frac{\pi}{2}
- \varphi$ is the azimuthal angle of the pion w.r.t.\ the reaction plane.
They are related to the ones previously used  by $q^2 \approx - \ft12
{\cal Q}^2$ and $\xi \approx \Bx/(2 - \Bx)$.

A calculation of the matrix element squared results into
\begin{eqnarray}
\label{XsectionFinal}
\frac{d \sigma^{\pi^+}}{d {\cal Q}^2 d \Bx d |\Delta^2| d \varphi}
=
\frac{\alpha^2_{\rm em}}{{\cal Q}^8}
\frac{f_\pi^2}{N_c^2} \frac{\Bx (1 - y)}{(2 - \Bx)^2}
\Bigg\{\!\!\!\!\!\!\!&&\!\!\!
8 (1 - \Bx) \widetilde {\cal H}^{ud \, \ast} \widetilde {\cal H}^{ud}
-
\frac{\Delta^2}{2 M^2} \Bx^2
\widetilde {\cal E}^{ud \, \ast} \widetilde {\cal E}^{ud}
-
4 \Bx^2 {\rm Re}
\left(
\widetilde {\cal H}^{ud \, \ast} \widetilde {\cal E}^{ud}
\right) \nonumber\\
&&\!\!\!\!\!\!\!-
4 \Bx \sqrt{1 - \Bx} \sqrt{- \frac{\Delta^2}{M^2}}
\sqrt{1 - \frac{\Delta^2_{\rm min}}{\Delta^2}} \sin \varphi \
{\rm Im}
\left(
\widetilde {\cal H}^{ud \, \ast} \widetilde {\cal E}^{ud}
\right)
\Bigg\},
\end{eqnarray}
where we have chosen the transverse polarization for the target proton $S =
(0, \cos {\mit\Phi}, \sin {\mit\Phi}, 0)$ along $y$-axis, ${\mit\Phi} =
\pi/2$, so that it does not possess the longitudinal component in the
laboratory frame with $z$-axis along the lepton beam. The azimuthal angle
$\varphi$ is between the proton spin and the projection of the pion momentum
on the $(x,y)$-plane. For the minimal momentum transfer we use the
approximation $\Delta^2_{\rm min} \approx - M^2 \Bx^2/(1 - \Bx)$, which is
valid for $\Bx M^2/{\cal Q}^2 \ll 1-\Bx$. We also dropped kinematical $\Bx^2
M^2/{\cal Q}^2$ corrections in the cross section (\ref{XsectionFinal}). The
functions $\widetilde {\cal H}$ and $\widetilde {\cal E}$ are introduced as
follows
\begin{eqnarray}
\label{MesonProFunctions}
\left\{
\begin{array}{c}
\widetilde {\cal H}^{ud}
\\
\widetilde {\cal E}^{ud}
\end{array}
\right\}
(\xi, \Delta^2)
= \int_0^1 d u \int_{-1}^{1} d x \
\phi_\pi (u)
T_{ud} \left( u, x, \xi \right)
\left\{
\begin{array}{c}
\widetilde H^{ud}
\\
\widetilde E^{ud}
\end{array}
\right\}
\left( x, \xi, \Delta^2 \right).
\end{eqnarray}

The conversion from leptoproduction to photoproduction, $d
\sigma_{\scriptscriptstyle \! L}^{\pi^+}$, with longitudinally polarized
photons $\varepsilon_{\scriptscriptstyle \! L}$ is done by multiplication
of the result (\ref{XsectionFinal}) by kinematical factors, namely,
\begin{equation}
\label{Conversion}
d \sigma_{\scriptscriptstyle \! L}^{\pi^+} = d \sigma^{\pi^+}
\left(
\frac{\left|
\varepsilon_{\scriptscriptstyle L} \cdot j
\right|^2}{q_1 \cdot P_1 }
\right)
\left(
\frac{\left| L \cdot j \right|^2}{k \cdot P_1 }
\frac{d^3 k'}{2 E' (2 \pi)^3}
\right)^{-1}
=
d \sigma^{\pi^+} \frac{1}{\alpha_{\rm em}}
\frac{\pi}{1 - y} \frac{\Bx}{d \Bx} \frac{{\cal Q}^2}{d {\cal Q}^2} .
\end{equation}
Here, to get the last equality we used $\left|
\varepsilon_{\scriptscriptstyle \! L} \cdot j \right|^2 = 4 {\cal Q}^2$,
$\left| L \cdot j \right|^2 = 16 (1 - y)/y^2$, $d^3 k'/( 2 E' ) = \pi
y/(2 \Bx) \, d \Bx d {\cal Q}^2$ and the known definition of $y$ resulting
from the ratio of the flux factors.

%%%%%%%%%%%%%%%%%%%%%%%%%%%%%%%%%%%%%%%%%%%%%%%%%%%%%%%%%%%%%%%%%%%%%
\section{Next-to-leading order corrections.}
\label{NLOanalytical}
%%%%%%%%%%%%%%%%%%%%%%%%%%%%%%%%%%%%%%%%%%%%%%%%%%%%%%%%%%%%%%%%%%%%%

Let us turn to the evaluation of the NLO corrections to the hard-scattering
amplitude. As we noted above in section \ref{HadronicAmplitude}, the
hard-scattering amplitude $T_{ud}$ can be deduced from that one computed in the
perturbative approach to the pion form factor\footnote{One has to use those
results where the symmetry properties of the distribution amplitudes, the
hard-scattering subprocess is convoluted with, was not used.} by the
replacement $v \to \frac{x + \xi}{2 \xi}$ and restoration of the imaginary
part according to the causal Feynman prescription. Thus, we decompose the
amplitude as
\begin{eqnarray}
\label{MesPro-TH}
&&\!\!\!\!T_{u d}
\left( u, x, \xi \left| {\cal Q}, \mu_F, \mu_R \right) \right.
\\
&&= C_F \alpha_s (\mu_R) \frac{1}{2 \xi}
\Bigg\{
Q_u T
\left( \left.
\frac{\xi - x}{2\xi} - i 0 , 1 - u \right|
{\cal Q}, \mu_F, \mu_R \right)
-
Q_d T
\left( \left.
\frac{\xi + x}{2\xi} - i 0 , u \right|
{\cal Q}, \mu_F, \mu_R \right)
\Bigg\},
\nonumber
\end{eqnarray}
where
\begin{eqnarray}
T \left( v, u \left| {\cal Q}, \mu_F, \mu_R \right)\right.
= \frac{1}{v u}
\left\{
1 + \frac{\alpha_s (\mu_R)}{2\pi}
T^{(1)} \left( v, u \left| {\cal Q}, \mu_F, \mu_R \right)\right.
+ {\cal O} \left( \alpha_s^2 \right)
\right\}.
\end{eqnarray}
Note that due to this $i 0$-prescription, $T_{u d}$ satisfy an unsubtracted
dispersion relation that is consistent with that one for the
$\gamma_{\scriptscriptstyle \! L} p \to \pi^+ n$ amplitude \cite{ColFraStr96}. 
Analogous to the pion form factor we extracted the LO coefficient function 
from the NLO amplitude $T^{(1)}$. In our consequent presentation we will drop 
for simplicity the dimensionfull arguments, i.e.\ ${\cal Q}$, the factorization 
scale $\mu_F$ and the renormalization scale $\mu_R$.

The one-loop correction $T^{(1)}$ for the pion form factor has been
calculated by different authors in dimensional regularization, however,
different renormalization and factorization prescriptions were applied
\cite{FieGupOttCha81,DitRad81,Sar82,RadKha85,BraTse87,MelNizPas98}. The
occurring differences were clarified in Refs.\ \cite{RadKha85,BraTse87}.
Indeed, if one takes into account errors, which are pointed out in Ref.\
\cite{BraTse87}, the results given in Refs.\ \cite{DitRad81,BraTse87} can be
transformed to those in Refs.\ \cite{FieGupOttCha81,Sar82,RadKha85,MelNizPas98}. 
They are given in the \MS\ scheme and the subtractions of infinities are done 
in a way that respects the universality of the distribution amplitude and the 
running coupling in this scheme. Thus, we have
\begin{equation}
\label{piofor-THNLO}
T^{(1)} = C_F T^F + \beta_0 T^\beta + \left(C_F - \frac{C_A}{2} \right) T^{FA} ,
\end{equation}
with
\begin{eqnarray}
\label{piofor-THNLOcoe}
T^F \!\!\!&=&\!\!\!
\left[ 3 + \ln(v u) \right] \ln\left( \frac{{\cal Q}^2}{\mu^2_F} \right)
+
\frac{1}{2} \ln^2(v u) + 3\ln(v u)
-
\frac{\ln v}{2(1 - v)}- \frac{\ln u}{2(1 - u)}
-
\frac{14}{3},
\nonumber\\
T^\beta \!\!\!&=&\!\!\!
\frac{1}{2}\ln\left( \frac{{\cal Q}^2}{\mu_R^2} \right) + \frac{1}{2} \ln(v u)
-
\frac{5}{6},
\nonumber\\
T^{FA} \!\!\!&=&\!\!\!
{\rm Li}_2 (1 - v) - {\rm Li}_2 (v)
+
\ln(1 - v) \ln\left( \frac{u}{1 - u} \right) - \frac{5}{3}
\nonumber \\
&+&\!\!\!
\frac{1}{(v - u)^2}
\Bigg\{
(v + u - 2 v u)\ln(1 - v) + 2 v u \ln(v)
\nonumber\\
&+&\!\!\!
\frac{(1 - v) v^2 + (1 - u) u^2}{v - u}
\left[ \ln(1 - v) \ln(u) - {\rm Li}_2 (1 - v) + {\rm Li}_2 (v) \right]
\Bigg\} + \{ v \leftrightarrow u \} ,
\end{eqnarray}
where ${\rm Li}_2 (u)= - \int_0^u \frac{d x}{x} \ln(1 - x)$  is the Euler
dilogarithm. The colour factors are conventionally defined by $C_A = N_c$,
$C_F = \frac{N_c^2 - 1}{2 N_c}$ and the first coefficient of the QCD beta
function reads $\beta_0 = \frac{4}{3} T_F N_f - \frac{11}{3} N_c$.

%%%%%%%%%%%%%%%%%%%%%%%%%%%%%%%%%%%%%%%%%%%%%%%%%%%%%%%%%%%%%%%%%%%%%
\section{Leading vs. next-to-leading order observables.}
\label{ScaleSet}
%%%%%%%%%%%%%%%%%%%%%%%%%%%%%%%%%%%%%%%%%%%%%%%%%%%%%%%%%%%%%%%%%%%%%

To provide estimates for the cross section and the transverse single spin
asymmetry as well as to discuss the scale setting ambiguities we use the
following models for the GPDs. For $\widetilde H^{ud}$ we assume a
factorizable ansatz of $(x, \xi)$ and $\Delta^2$ dependence valid at small
$\Delta^2$. In terms of the double distribution $F (y, z)$, we have
\begin{equation}
\label{Htilde}
\widetilde H^{ud} (x, \xi, \Delta^2)
= G (\Delta^2)
\int_{-1}^{1} d y \int_{- 1 + |y|}^{1 - |y|} d z
\delta (y + \xi z - x) F^{ud} (y, z) .
\end{equation}
Here the form factor has a dipole parametrization $G (\Delta^2)
= (1 - \Delta^2/m_A^2)^{-2}$ with $m_A^2 = 0.9 \ {\rm GeV}^2$ and unit
normalization. The double distribution is modeled according to Ref.\
\cite{Rad99}
\begin{equation}
\label{RadMod}
F (y, z)
= \left\{
\Delta q (y) \theta (y) - \Delta \bar q (- y) \theta (- y)
\right\} \pi (|y|, z) ,
\qquad
\pi (y, z) = \frac{3}{4} \frac{(1 - y)^2 - z^2}{(1 - y)^3} .
\end{equation}
Assuming an $SU (2)$ symmetric sea, i.e.\ $\Delta \bar q^{ud} (y) = \Delta \bar u
(y) - \Delta \bar d (y) = 0$, we have $\Delta q^{ud} (y) = \Delta u_{\rm val}
(y) - \Delta d_{\rm val} (y)$. For the evaluations given below we use GSA
forward parton densities at $Q^2 = 4 \ {\rm GeV}^2$ \cite{GehSti96}.
Note that the simple ansatz (\ref{Htilde}) for $\widetilde H^{ud}$ is chosen
in such a way that constraints arising from the reduction to the forward
limit and the sum rule for the lowest moment are satisfied.

For $\widetilde E^{ud}$ we adopt the pion pole-dominated ansatz 
\begin{equation}
\label{Etilde}
\widetilde E^{ud} (x, \xi, \Delta^2) = F_\pi (\Delta^2)
\frac{\theta \left( \xi > |x| \right)}{2\xi}
\phi_\pi\!\!\left(\frac{x+\xi}{2\xi}\right) ,
\end{equation}
with $F_\pi (\Delta^2)$ taken in our estimates in the form given in Eq.\
(39) of Ref.\ \cite{PenPolGoe00}. In the vicinity of the pion pole this form
factor reads $F_\pi ( \Delta^2 \to m_\pi^2 ) = 4 g_A M^2 / \left( m_\pi^2 -
\Delta^2 \right)$. For the pion distribution amplitude $\phi_\pi$ we will
take for simplicity its asymptotic form\footnote{It results into a good
agreement of the theoretical predictions for the photon-to-pion transition
form factor with experimental data. However, one should be aware that other
quite different distribution amplitudes are consistent with the data
(see for instance \cite{BakMikSte01}).}
\begin{eqnarray}
\phi_\pi (u) = \phi^{\rm asy}(u) \equiv 6 u (1 - u),
\qquad
0 \leq u \leq 1 .
\end{eqnarray}

In the following we will discard the evolution effects of the GPDs for
simplicity since they are small for the models and initial scale we chose.
Especially, they are negligible for the asymptotic pion distribution
amplitude since they arise solely from two-loop (and higher) effects.
Same applies to $\widetilde E$ since it is proportional to the asymptotic
distribution amplitude, too, i.e.\ $\xi \widetilde E^{ud} (x, \xi) \propto
\phi_\pi \left( ( x + \xi )/ (2 \xi) \right)$. The neglected effects are
of order $1 \%$ or so \cite{BelMul98}. For $\widetilde H$ the evolution
is more prominent since it shows up already at LO. However, it is still small for
the change of scale from $4 \ {\rm GeV}^2$ to $10 \ {\rm GeV}^2$ and result
in a $5 - 8 \%$ change in the shape \cite{BelMul98}.

Let us now discuss the scale setting issues in the NLO coefficient function.
A truncation of the perturbative series at a given order of coupling, here
at NLO, causes a residual dependence on the factorization and
renormalization scales. Obviously, this fact implies scale setting
ambiguities, which result into uncertainties for the theoretical
predictions. There are different suggestions to solve the scale setting
problem with the aim to minimize the unknown higher order corrections.
Unfortunately, they can provide quite different theoretical predictions for
observables.

We start with the discussion of the $\mu_F$ dependence, which is intimately
related to the evolution of the distribution amplitudes/GPDs. Note that this
dependence completely disappears in $\widetilde {\cal E}$, if we assume (as
we do) that $\phi_\pi$ coincides with the asymptotic distribution amplitude
and one neglects its NLO evolution. The coefficient of $\ln\left({\cal Q}^2/{\mu^2_F}
\right)$ in Eq.\ (\ref{piofor-THNLOcoe}) is merely the LO evolution
kernel (convoluted with the LO coefficient function) and the asymptotic
distribution amplitude is its eigenfunction with zero eigenvalue, e.g.\
\begin{eqnarray*}
\int_{-1}^{1} dx \ \widetilde E^{ud} (x, \xi)
\left( 3 + 2 \ln \frac{x \pm \xi}{2 \xi} \right) = 0 .
\end{eqnarray*}
In $\widetilde {\cal H}$ the term proportional to $\ln\left({\cal
Q}^2/{\mu^2_F} \right)$ survives now in the convolution with $\widetilde H$.
However, as we pointed out above the evolution at LO is weak and the NLO
corrections are small \cite{BelMul98}. This implies that the
$\mu_F$-dependence in the $\widetilde {\cal H}$ amplitude is very weak.
Therefore, we simply set $\mu_F = {\cal Q}$ in what follows. A detailed
discussion of other plausible settings for $\mu_F$ runs beyond the scope
of the present paper.

Next we turn to the $\mu_R$ scale. The variation of the amplitude with this
scale appears at LO from the change of $\alpha_s(\mu_R)$. At NLO this scale
ambiguity is expected to be smaller since this change is partially canceled 
by the variation of the term proportional to $\beta_0$ in Eq.\
(\ref{piofor-THNLOcoe}). For reasons explained below, we will concentrate
here only on two possibilities:
\begin{itemize}
\item The renormalization scale is set equal ${\cal Q}$, $\mu_R = {\cal Q}$.
\item The Brodsky-Lepage-Mackenzie (BLM) scale setting
prescription \cite{BroLepMak83}.
\end{itemize}

For the naive setting we observe in both $\widetilde{\cal H}$, separately
for its real and imaginary part, and in $\widetilde{\cal E}$ function a
cancelation of two large contributions: a positive one proportional to $(-
\beta_0)$ and a negative one proportional to $C_F$. The term proportional to
$( C_F - C_A/2 )$ in Eq.\ (\ref{piofor-THNLO}) is not numerically important
since it is relatively suppressed by $1/N_c^2$. Taken together they result
for $N_f = 3$ into a large positive effect.

The naive scale settings may not be quite appropriate by different reasons.
The renormalization and factorization scales reflect quite different types of
perturbative corrections \footnote{In the consequent discussion we follow
analogous considerations for the pion form factor in Ref.\ \cite{BakRadSte00}}.
The first one arises from the renormalization of
ultraviolet divergences and generates the running of $\alpha_s$, while the
second one comes from the factorization of collinear singularities and
provides the evolution of the distribution amplitudes and GPDs. The former
originates from the geometric series with expansion terms having definite
sign. On the other hand, the latter is of the evolution and remnant Sudakov
type effects \cite{BakRadSte00}. The exponentiated Sudakov corrections have 
a sign alternating expansion. Thus, while both partially cancel at NLO, it is
expected that they will amplify at NNLO. Furthermore, since the hard
external scattering scale is partitioned among a number of parton
participants in the hard-scattering, their resulting virtuality is much
softer than the external hard scale. Thus, the coupling in the quark-gluon
emission vertex has to be taken at a mean virtuality of e.g.\ hard gluon. A
generalization of this idea results into a sensible prescription to absorb
all effects coming from the running of the coupling, i.e.\ terms
proportional to the $\beta$-function, into the scale of the coupling. This
is the BLM procedure \cite{BroLepMak83}. Since the trace anomaly of the
energy momentum tensor is proportional to the $\beta$-function, the
resulting series in $\alpha_s$, if using this procedure, formally coincides
with the perturbation theory in the conformally invariant (massless) QCD.

Since in general we have to deal with two different amplitudes defined in
Eq.\ (\ref{MesonProFunctions}), which contain a real and imaginary part, the
consequent application of the BLM setting procedure requires the
introduction of different scales. We separately determine the scales for the real
and imaginary parts of the functions $\widetilde {\cal H}$ and $\widetilde
{\cal E}$ from the conditions
\begin{equation}
\int_0^1 \frac{d u}{u}
\int_{-1}^{1} d x \ 
\phi_\pi (u)
\left\{
Q_u \frac{T^\beta \left( \frac{\xi - x}{2 \xi} - i 0, u\right)}{\xi - x - i 0}
-
Q_d \frac{T^\beta \left( \frac{\xi + x}{2 \xi} - i 0, u\right)}{\xi + x - i 0}
\right\}
\left\{
\begin{array}{c}
\widetilde H^{ud}
\\
\widetilde E^{ud}
\end{array}
\right\}
\left( x, \xi, \Delta^2 \right)
= 0.
\end{equation}
Note that the scales will depend on the shape of the distribution amplitude and
the GPDs.

For $\widetilde {\cal E}$ there is no imaginary part since $\widetilde E$, as
defined in Eq.\ (\ref{Etilde}), is concentrated solely in the exclusive
domain. For its real part due to $x \to -x$ symmetry both the $Q_u$ and
$Q_d$ contributions in Eq.\ (\ref{MesPro-TH}) are the same and one immediately
finds that the BLM scale
\begin{equation}
\mu_R^2 = {\cal Q}^2 {\rm e}^{-14/3} 
\end{equation}
coincides with the one of the pion form factor \cite{BroJiPanRob98,BakRadSte00}
for the asymptotic distribution amplitude.

To discuss the scale setting in $\widetilde {\cal H}$, let us first consider
the properties of $\widetilde H$. From Eqs.\ (\ref{Htilde},\ref{RadMod})
with $\Delta \bar q = 0$ it is obvious that the function vanishes for $x < -
\xi$, $\widetilde H^{ud} ( x < - \xi, \xi) = 0$. From this we conclude that
the imaginary part of $Q_d$ contribution vanishes. As we mentioned
above, for $\widetilde H$ the scale will be different for the imaginary
and real parts and will depend on the skewedness
\begin{equation}
\mu_R^2 = {\cal Q}^2 {\rm e}^{- f ( \xi)} .
\end{equation}
For the imaginary part one gets
\begin{equation}
f_{\rm Im} (\xi) = \frac{19}{6} - \ln \frac{1 - \xi}{2 \xi}
- \int_{\xi}^1 d x \ \frac{1 - \widetilde H^{ud} (x, \xi) /
\widetilde H^{ud} (\xi, \xi)}{\xi - x} .
\end{equation}
The scale for the real part of the $\widetilde {\cal H}$-contribution is 
governed by the function
\begin{equation}
\label{Real-u}
f_{{\rm Re}} (\xi)
= \frac{19}{6}
- \frac{
\int_{-1}^{1} d x \
\left\{
Q_u \frac{{\rm PV}}{\xi - x} \ln \frac{| \xi - x |}{2 \xi}
-
Q_d \frac{{\rm PV}}{\xi + x} \ln \frac{| \xi + x |}{2 \xi}
+ \frac{\pi^2}{2}
\left(
Q_u \delta (\xi - x)
-
Q_d \delta (\xi + x)
\right)
\right\}
\widetilde H^{ud} (x, \xi)}
{\int_{-1}^{1} d x \
\left\{
Q_u \frac{{\rm PV}}{\xi - x}
-
Q_d \frac{{\rm PV}}{\xi + x}
\right\}
\widetilde H^{ud} (x, \xi)},
\end{equation}
where the symbol ${\rm PV}$ stands here for the Cauchy principal value
prescription.

For the models we are using in all of the three cases ${\rm e}^{-f}$ is
below $0.1$ in a wide interval of Bjorken variable $0.1 < \Bx < 0.5$. This
means that the argument of the coupling is driven into the infrared region.
There are experimental indications that the coupling is frozen at such a low
scale \cite{Fro80}. Indeed, in exclusive processes, such as nucleon form
factors, fixed angle proton-proton elastic scattering etc., the data at
large scales are consistent with the predictions of dimensional counting
rules. On the other hand, the perturbative leading twist QCD analyses
coincides with the dimensional counting rules, however, is proportional to powers
of $\alpha_s$, e.g.\ $\alpha_s^2 / Q^4$ and $\alpha_s^6 / t^8$,
respectively. Since higher order analyses favour a low scale in the
coupling, we may conclude that the latter is a slowly varying function in
this domain \cite{Fro80}. For our purposes we choose a frozen coupling with
$\alpha_s (\mu_R^2)/\pi = 0.1$ for $\mu_R^2 < 1 \ {\rm GeV}^2$
\cite{BroJiPanRob98,BakRadSte00} instead of using three different soft
scales in the analytical coupling, e.g.\ with gluon mass $\mu_R^2 \to
\mu_R^2 + 4 m_g^2$.

Now we are in a position to present numerical estimates of the observables
at LO and NLO. Due to the asymptotic form of the distribution amplitude,
which we use, the integration w.r.t.\ the momentum fraction $u$ can be done
analytically:
\begin{eqnarray}
{\cal T}_{ud} (x, \xi)
\!\!\!&=&\!\!\!
\int_0^1 d u \, \phi^{\rm asy} (u) T_{ud} (u, x, \xi)
\\
\!\!\!&=&\!\!\!
C_F \alpha_s (\mu_R) \frac{1}{2 \xi}
\left\{
Q_u {\cal T}
\left(
\frac{\xi - x}{2\xi} - i0
\right)
-
Q_d {\cal T}
\left(
\frac{\xi + x}{2\xi} - i 0
\right)
\right\} ,
\nonumber
\end{eqnarray}
where
\begin{eqnarray}
{\cal T} (v) \!\!\!&=&\!\!\!
\int_0^1 d u \, \phi^{\rm asy} (u) T \left( v, u \right)
\\
&\!\!\!=\!\!\!&
\frac{1}{v}
\left\{
3
+
\frac{\alpha_s (\mu_R)}{2 \pi}
\left(
C_F {\cal T}^F (v)
+
\beta_0 {\cal T}^\beta (v)
+
\left( C_F - \frac{C_A}{2} \right)
{\cal T}^{FA} (v)
\right)
+
{\cal O} \left( \alpha_s^2 \right)
\right\} ,
\nonumber
\end{eqnarray}
with
\begin{eqnarray}
{\cal T}^F \!\!\!&=&\!\!\!
\frac{3}{2} \left[ 3 + 2\ln v \right] \ln\left( \frac{{\cal Q}^2}{\mu^2_F} \right)
+
\frac{3}{2} \ln v \left[ 3 + \ln v \right]
-
\frac{3 \ln v}{2(1 - v)} - \frac{77}{4} ,
\nonumber\\
{\cal T}^\beta \!\!\!&=&\!\!\!
\frac{3}{2} \ln\left( \frac{{\cal Q}^2}{\mu_R^2} \right) + \frac{3}{2} \ln v
-
\frac{19}{4},
\nonumber\\
{\cal T}^{FA} \!\!\!&=&\!\!\!
- 1 - 6 [ \ln v + 2 \ln (1 - v) ]
+
12 (1 - 3 v) \left\{ {\rm Li}_2 (v) - {\rm Li}_2 (1 - v)
+ \ln \frac{v}{1 - v} \right\}
\nonumber\\
&&\!\!\!+ 6 (1 - 6 v + 6 v^2)
\Bigg\{
3 {\rm Li}_3 (v) + 3 {\rm Li}_3 (1 - v)
- \ln\frac{v}{1 - v} \, {\rm Li}_2 (v)
+ \ln^2 (1 - v) \ln v
\nonumber\\
&&\qquad\qquad\qquad\qquad
- \frac{\pi^2}{6} [\ln v + 2 \ln (1 - v) ]  \Bigg\},
\end{eqnarray}
and  ${\rm Li}_3(v)= \int_{0}^v d u {\rm Li}_2(u)/u $.

\begin{figure}[t]
\vspace{-1cm}
\begin{center}
\mbox{
\begin{picture}(0,310)(300,0)
\put(70,150){\insertfig{8}{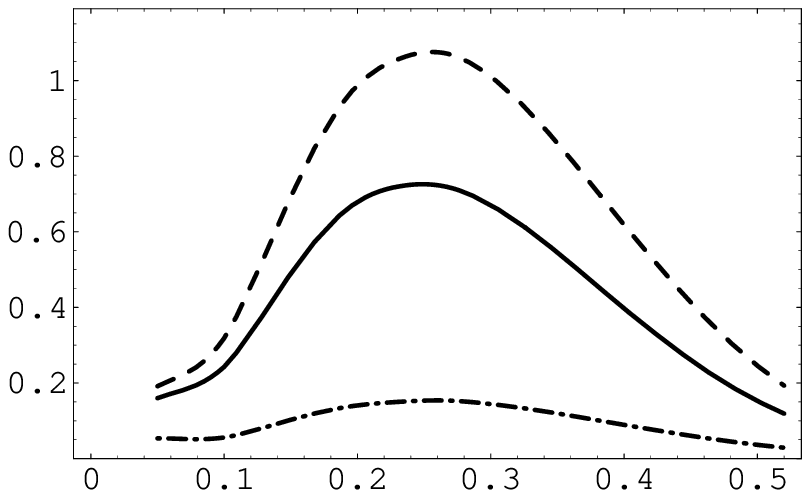}}
\put(273,270){(a)}
\put(50,165){\rotate{$d \sigma^{\pi^+}_{\scriptscriptstyle \! L} \! /
d |\Delta^2| \, ({\rm nb}/{\rm GeV}^2)$}}
\put(310,150){\insertfig{8}{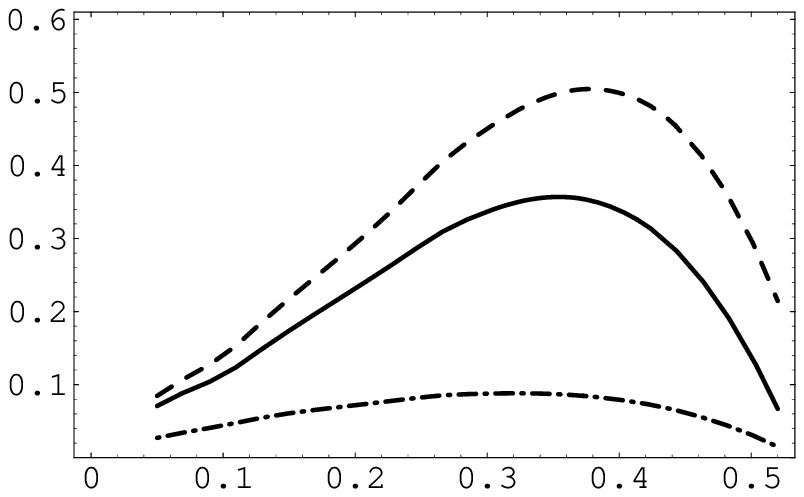}}
\put(512,270){(b)}
\put(70,0){\insertfig{8}{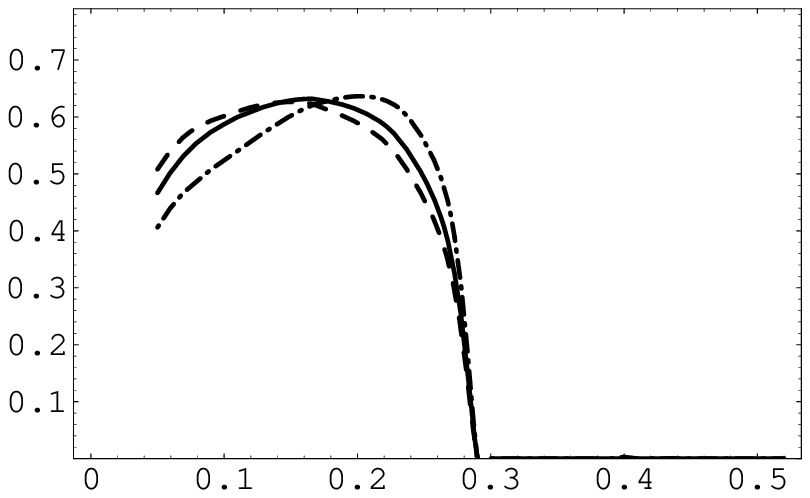}}
\put(273,118){(c)}
\put(50,65){\rotate{$A_\perp$}}
\put(280,-10){$x_{\rm B}$}
\put(310,0){\insertfig{8}{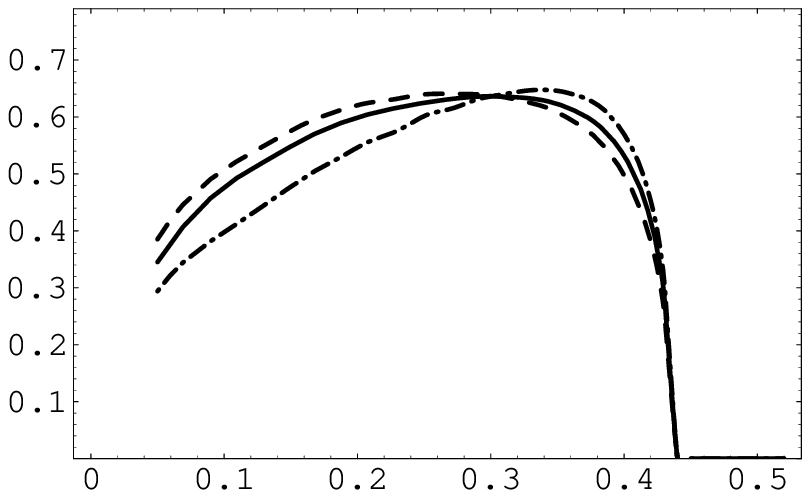}}
\put(512,118){(d)}
\put(520,-10){$x_{\rm B}$}
\end{picture}
}
\end{center}
\caption{\label{Fig-Pred} The leading twist predictions for the unpolarized
photoproduction cross section $d\sigma_{\scriptscriptstyle \! L}^{\pi^+}/d |
\Delta^2 |$ at ${\cal Q} = 10 \ {\rm GeV}^2$ for the GPD models specified 
in Eqs.\ (\ref{Htilde},\ref{Etilde})
are shown for $\Delta^2 = \Delta^2_{\rm min}$ and $\Delta^2 = - 0.3 \
{\rm GeV}^2$ in the panels (a) and (b), respectively. In (c) and (d) we
display the transverse proton single spin asymmetry $A_\perp$ for $\Delta^2 
= - 0.1 \ {\rm GeV}^2$ and $\Delta^2 = - 0.3 \ {\rm GeV}^2$, respectively. 
The solid, dashed and dash-dotted curves represent the LO and NLO with the 
naive and BLM scale setting, respectively.}
\end{figure}

The LO predictions were done for the running coupling with $\mu_R = {\cal Q}$,
$\Lambda^{\rm LO}_{\rm QCD} = 220 \ {\rm MeV}$ and $N_f = 3$. At NLO we use,
as discussed above, two scale setting procedures: the naive $\mu_R = {\cal Q}$
and BLM one, with the running $\alpha_s ({\cal Q}^2)$ in the first and a fixed
$\alpha_s/\pi = 0.1$ below $1 \ {\rm GeV}^2$ in the second case, respectively.
Our LO predictions given in Fig.\ \ref{Fig-Pred} (a, b) are in agreement with
recent analyses in Refs.\ \cite{ManPilRad99,FraPobPolStr99,VanGuiGui99}. However,
they are plagued by large uncertainties from the higher order corrections. It is
interesting to note that taking forward parton distribution as a model for
the GPD, one gets almost identical results for the cross section.

The transverse single spin asymmetry defined by
\begin{equation}
\label{tSSA}
A_\perp
= \left(
\int_{0}^{\pi}
d \varphi \frac{d \sigma_{\scriptscriptstyle \! L}^{\pi^+}}{d |\Delta^2| d \varphi}
-
\int_{\pi}^{2 \pi}
d \varphi \frac{d \sigma_{\scriptscriptstyle \! L}^{\pi^+}}{d |\Delta^2| d \varphi}
\right)
\left(
\int_{0}^{2 \pi}
d \varphi \frac{d \sigma_{\scriptscriptstyle \! L}^{\pi^+}}{d |\Delta^2| d \varphi}
\right)^{- 1} 
\end{equation}
is studied numerically in Fig.\ \ref{Fig-Pred} (c, d) for $\Delta^2 =
- (0.1,\ 0.3) \ {\rm GeV}^2$. Our leading predictions shows the
same large asymmetry as has originally been observed in Ref.\
\cite{FraPobPolStr99}. As our studies demonstrate this observable is less
sensitive to higher order effects.

%%%%%%%%%%%%%%%%%%%%%%%%%%%%%%%%%%%%%%%%%%%%%%%%%%%%%%%%%%%%%%%%%%%%%
%            Figure 4
%%%%%%%%%%%%%%%%%%%%%%%%%%%%%%%%%%%%%%%%%%%%%%%%%%%%%%%%%%%%%%%%%%%%%

%%%%%%%%%%%%%%%%%%%%%%%%%%%%%%%%%%%%%%%%%%%%%%%%%%%%%%%%%%%%%%%%%%%%%
\section{Discussion and conclusions.}
%%%%%%%%%%%%%%%%%%%%%%%%%%%%%%%%%%%%%%%%%%%%%%%%%%%%%%%%%%%%%%%%%%%%%

We have discussed in the present Letter the cross section and transverse
proton single spin asymmetry in the hard exclusive production of pions. For
the longitudinal polarization of the virtual photon the amplitude is
dominated by twist-two contributions, i.e.\ lowest Fock components in the
pion and $p \to n$ transition amplitudes. We evaluated the NLO corrections
to the hard parton subprocess basing on the available result for the pion
form factor and estimated their size in physical observables. We found that
the ambiguity in the renormalization scale setting (together with the
freezing of the coupling in the infrared region) results into large ${}^{+
40 \%}_{- 70 \%}$ uncertainties in the theoretical predictions, $(T_{NLO} -
T_{LO})/T_{LO}$, of the amplitude. After application of the BLM scale
setting prescription, we observed a large reduction in the magnitude of 
the cross section related to the effects of Sudakov double logarithms. A 
deeper insight into the structure of these corrections is highly desirable.

The theoretical uncertainty in the factorization procedure on the
amplitude level is translated into large variations of the physical cross
section. However, we found that the single spin asymmetry, given by the ratio
of the Fourier coefficients of the cross section, is a `good' observable
since the ambiguities due to the truncation of the perturbative series
approximately cancel. Thus, the perturbative predictions for this quantity
are rather stable. The NLO effects result into ${}^{+ 7 \%}_{- 18 \%}$
corrections to the LO prediction for $0.1 < \Bx < 0.5$. For the assumed
models of the non-perturbative functions the asymmetry is big, being as
large as $60 \%$, and is sensitive to the pion pole dominated GPD $\widetilde
E$. Thus, in view of its advantages being rather insensitive to the higher
order corrections, it turns out to be an appropriate quantity for
experimental studies at Jefferson Lab and HERMES. The large NLO
corrections may indicate that the application of the perturbative approach
to the meson production cross section is legitimate at a rather large
momentum transfer. However, for the asymmetry, which depends at leading
twist only logarithmically on ${\cal Q}^2$, one expects its earlier validity
due to an essential cancelation of higher order perturbative and, hopefully,
also power suppressed corrections, due to their intrinsic interrelation in
field theoretical treatment, recall renormalons in this respect.

Let us stress that the longitudinal proton single spin asymmetry, measured at
HERMES \cite{Ste01}, arises at the twist-three level and requires a
separate analysis. The result we have presented here are given for the
transversely polarized proton target. The NLO analysis addressed in our
study can be extended to a large set of meson leptoproduction amplitudes
with the quark dominated short-distance subprocess, e.g.\ $\ell p \to \ell'
\pi^0 p$, $\ell p \to \ell' \rho^+ n$, etc.

\vspace{0.5 cm}

We would like to thank J.C. Collins, P. Kroll, J. Smith, M. Stratmann for 
useful conversations and especially A.V. Radyushkin for illuminating and 
instructive discussions. D.M. is grateful to J. Smith for the hospitality 
extended to him at the C.N. Yang Institute for Theoretical Physics where 
a major part of this work has been written.

\end{document}